\documentclass[preprint,aps,pra,groupedaddress,amsmath,amssymb,showpacs]{revtex4-1}
\usepackage{graphicx,amsmath,amssymb,epsfig}
\usepackage{dcolumn}
\def\be{\begin{equation}}
\def\ee{\end{equation}}
\def\bea{\begin{eqnarray}}
\def\eea{\end{eqnarray}}
\def\bes{\begin{subequations}}
\def\ees{\end{subequations}}

\begin{document}

\title{Ultraslow Helical Optical Bullets and Their Acceleration in
Magneto-Optically Controlled Coherent Atomic Media}
\author{Chao Hang}
\affiliation{State Key Laboratory of Precision Spectroscopy and Department of Physics,
East China Normal University, Shanghai 200062, China}
\author{Guoxiang Huang}
\email[Corresponding author: ]{gxhuang@phys.ecnu.edu.cn}
\affiliation{State Key Laboratory of Precision Spectroscopy and
Department of Physics, East China Normal University, Shanghai
200062, China}
\date{\today}

\begin{abstract}

We propose a scheme to produce ultraslow (3+1)-dimensional helical
optical solitons, alias helical optical bullets, in a
resonant three-level $\Lambda$-type atomic system via
quantum coherence. We show that, due to the effect of
electromagnetically induced transparency, the helical optical
bullets can propagate with an ultraslow velocity up to
$10^{-5}$ $c$ ($c$ is the light speed in vacuum) in longitudinal
direction and a slow rotational motion (with velocity
$10^{-7}$ $c$) in transverse directions. The generation
power of such optical bullets can be lowered to
microwatt, and their stability can be achieved by using a Bessel optical
lattice potential formed by a far-detuned laser field. We also show
that the transverse rotational motion of the optical
bullets can be accelerated by applying a time-dependent Stern-Gerlach
magnetic field. Because of the untraslow velocity in the longitudinal direction,
a significant acceleration of the rotational motion of optical
bullets may be observed for a very short medium length.

\end{abstract}

\maketitle


\section{Introduction}


In recent years, the formation and propagation of a new type of optical solitons,
i.e. ultraslow optical solitons, created in resonant multi-level media via
electromagnetically induced transparency (EIT) \cite{fle}, has
attracted much attention \cite{wu,huang,hang1,yang}.
By the quantum interference effect induced by a control field,
the absorption of a probe field can be largely suppressed. Simultaneously,
a drastic change of dispersion and a giant enhancement of Kerr nonlinearity
can be achieved via EIT. It has been shown that,
by the balance of the dispersion and the Kerr nonlinearity,
temporal optical solitons with very small propagating velocity and very low
generation power can form and propagate stably for a long distance.
The possibility of producing weak-light spatial and spatia-temporal
optical solitons in EIT-based media has also been
suggested~\cite{hong,hang2,MPP,HK,LWH}.

On the other hand, the optical beam deflection in an atomic
system with inhomogeneous
external magnetic fields has been the subject of many previous
works~\cite{SW,hol,ZZWYS}. Physically, atomic energy levels are
space-dependent if an inhomogeneous magnetic field is applied,
which results in a spatial dependence of optical refraction index
of the atomic medium and hence a deflection of optical beam when
passing through the medium. In a remarkable experiment~\cite{KW},
Karpa and Weitz demonstrated that photons
can acquire effective magnetic moments when propagating in EIT
media, and hence can deflect significantly in a transverse
gradient magnetic field, which is regarded as an optical analog of
Stern-Gerlach (SG) effect of atoms. The work by Karpa and
Weitz~\cite{KW} has stimulated a flourished theoretical and
experimental studies on the SG effect of slow light via
EIT~\cite{ZLZS,liy,GZKS,KW1}. Recently, a scheme to exhibit
EIT-enhanced SG deflection of weak-light vector
optical solitons in an EIT medium has been suggested~\cite{HH}.

In this article, we propose a scheme to produce ultraslow
(3+1)-dimensional helical optical solitons, alias helical optical
bullets, in a resonant three-level $\Lambda$-type atomic
system via quantum coherence. We show that due to the EIT effect the
helical optical bullets found can propagate with ultraslow
velocity up to $10^{-5}$ $c$ ($c$ is the light speed in vacuum) in
longitudinal direction and a slow rotational motion (with
velocity  $10^{-7}$ $c$) in transverse directions. The
generation power of such optical bullets can be lowered to the
magnitude of microwatt, and their stability can be achieved by
using a Bessel optical lattice potential formed by a far-detuned
laser field. We also show that the transverse rotational
motion of the optical bullets can be accelerated by the use of a
time-dependent SG  magnetic field. Because of the ultraslow velocity
in the longitudinal direction, a significant acceleration of the
rotational motion of the optical bullets may be observed for
a very short medium length.

We stress that the ultraslow helical optical bullets found here
have many attractive features. First, because of the giant Kerr
nonlinearity induced by the EIT effect the optical bullets can
form in a very short distance (at the order of centimeter) with
extremely low generation power (at the order of microwatt).
Second, they have ultraslow propagating velocity which prolongs
the interaction time between light and atoms. Third, their
motional trajectories are helical curves (see Fig.~\ref{fig4}(c)
below). Fourth, because of the active character of our system it
is very easy to realize an efficient magneto-optical manipulation
of the helical optical bullets
by using the energy-level structure of atoms and selection
rules of optical transitions. Due to these features, the ultraslow
helical optical bullets obtained here may become candidates for
light information processing and transmission at a very weak light
level. Note that in recent years several authors have done
interesting works on linear and nonlinear matter waves
in Bose-Einstein condensates with toroidal traps and realized
their acceleration~\cite{JPY1,DJM,RACN,BK}. However, in those
works no ultraslow bullets and their helical motion have been
obtained.

Our article is arranged as follows. In Sec. II, the
physical model of a three-state atomic system with a $\Lambda$-type
energy-level configuration is described. Maxwell-Bloch (MB) equations governing
the motion of density matrix elements and the probe-field Rabi frequency
are given. In Sec. III, a nonlinear envelope equation governing the motion
of the probe-field is derived based on a standard method of multiple-scales.
In Sec. IV, the formation and propagation of ultraslow
high-dimensional helical optical bullets are discussed.
In Sec. V, the stability of helical optical solitons is
analyzed and their acceleration is investigated in detail. The last section
(Sec. VI) contains a discussion and summary of main results of our work.


\section{Model}

\begin{figure}[tbph]
\centering
\includegraphics[scale=0.4]{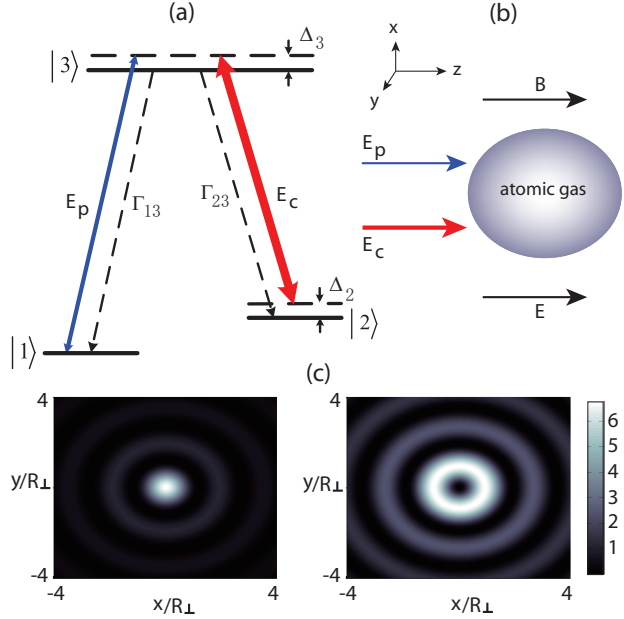}
\caption{{\protect\footnotesize (Color online) (a): Energy-level
diagram and excitation scheme of $\Lambda$-type three-level atoms
interacting with a weak pulsed probe field $E_{p}$ and a strong CW
control field $E_{c}$. $\Delta _{2}$ and $\Delta _{3}$ are the two-
and one-photon detunings, respectively.
$\Gamma_{13}$ and $\Gamma_{23}$ are the decay rates from
$|3\rangle$ to $|1\rangle$ and from $|3\rangle$ to $|2\rangle$,
respectively. (b): The coordinate frame used for calculations and
geometrical arrangement of the system. The shadow region denotes
the atomic gas confined in a cell,  $B$ and $E$
are gradient magnetic field and far-detuned laser field,
respectively. (c): Left (right) part corresponds to the light-intensity distribution
of the zero (first) order Bessel lattice $J_0^2(2r)$ ($J_1^2(2r)$\,)
in the $(x,y)$ plane, with $r=(x^2+y^2)^{1/2}$ and $R_{\bot}$
the probe beam radius. }}
\label{fig1}
\end{figure}

We consider a resonant, lifetime-broadened atomic gas with a $\Lambda$-type energy-level
configuration (Fig.~\ref{fig1}(a)\,), which
interacts with a strong, continuous-wave (CW)
control field of angular frequency $\omega _{c}$ that drives the
transition $|2\rangle \leftrightarrow |3\rangle $  and a weak,
pulsed probe field (with the pulse length $\tau _{0}$ and radius
$R_{\bot}$ at the entrance of the medium) of center angular
frequency $\omega _{p}$ that drives the transition $|1\rangle
\leftrightarrow |3\rangle$, respectively.
$\Delta _{2}$ and $\Delta _{3}$ are respectively the two-
and one-photon detunings,
$\Gamma_{13}$ and $\Gamma_{23}$ are respectively the decay rates
from $|3\rangle$ to $|1\rangle$ and from $|3\rangle$ to $|2\rangle$.
The electric-field vector of the system can be written as
$\mathbf{E}=\mathbf{E}_p+\mathbf{E}_c=\sum_{l=c,p}{\mathbf{e}_{l}\mathcal{E}}_{l}\exp
[i(k_l z-\omega_l t)]+{\rm c.c}.$, where $\mathbf{e}_{c} $ and
$\mathbf{e}_{p}$  ($\mathcal{E}_{c}$ and $\mathcal{E}_{p}$)
are  respectively the polarization unit vectors (envelopes) of the
control and probe fields, $k_p=\omega_p/c$ and $k_c=\omega_c/c$
are respectively the center wavenumbers of the probe and control fields.
For simplicity, both the probe and the control fields are taken to
propagate along $z$-direction.

We assume a weak, time-dependent gradient magnetic field with the form
\be \label{SGfield} {\bf B}(t)=\hat{\bf z}f(t)Bx, \ee
is applied to the system, where  $\hat{\bf z}$ is the
unit vector in $z$-direction and $B$ characterizes
magnitude of the gradient and $f(t)$ describes its time dependence.
Due to the presence of ${\bf B}(t)$, Zeeman level shift $\Delta E_{j,
\rm Zeeman}=\mu_B g_F^{j} m_F^{j} f(t)Bx$ occurs for all levels. Here $\mu_B$,
$g_F^{j}$, and $m_F^{j}$ are Bohr magneton, gyromagnetic factor,
and magnetic quantum number of the level $|j\rangle$,
respectively. The aim of introducing the SG gradient
magnetic field (\ref{SGfield}) is to produce an external force in transverse
directions to control the motion of optical bullets
formed by the probe field.

In addition, we assume further a weak,
far-detuned laser field with the form
\be \label{Efield} {\bf E}(r,t)=\hat{\bf
z}\,pJ_l(\sqrt{2b}r)\cos(\omega_E t), \ee
is also applied into the system. Here $r=\sqrt{x^2+y^2}$,
$J_l$ is the $l$-order Bessel function with $p$ and $b$
characterizing respectively its amplitude and radius,
and $\omega_E$ is oscillating angular frequency.
Due to the presence of ${\bf E}(r,t)$, Stark level shift $\Delta
E_{j,\rm Stark} =-\frac{1}{2}\alpha_j \langle
E^2\rangle_t=-\frac{1}{4}\alpha_j p^2 J_l^2(\sqrt{2b}r)$ occurs for all levels,
where $\alpha_j$ is the scalar polarizability of the level
$|j\rangle$. Shown in the left (right) part of Fig.~\ref{fig1}(c)
is the light-intensity distribution corresponding to the zero (first) order Bessel
lattice $J_0^2(2r)$ ($J_1^2(2r)$\,).
The aim of introducing the far-detuned laser field (\ref{Efield}) is
to form a trapping lattice potential in the transverse directions to stabilize
the optical bullets. A possible geometrical arrangement of the system
is given in Fig.~\ref{fig1}(b).

Under electric-dipole and rotating-wave approximations, in interaction picture equations
of motion for the density matrix elements are given by
\begin{subequations}
\label{BLO}
\begin{eqnarray}
& &i\frac{\partial }{\partial t}\sigma _{11}
-i\Gamma _{13}\sigma _{33}+\Omega _{p}^{\ast }\sigma _{31}-\Omega
_{p}\sigma _{31}^{\ast }=0,  \label{1a} \\
& &i\frac{\partial }{\partial t}\sigma _{22}
-i\Gamma _{23}\sigma _{33}+\Omega _{c}^{\ast }\sigma _{32}-\Omega
_{c}\sigma _{32}^{\ast }=0,  \label{1b} \\
& &i\frac{\partial }{\partial t}\sigma _{33}+i\Gamma _{3}\sigma _{33}-\Omega
_{p}^{\ast }\sigma _{31}+\Omega _{p}\sigma _{31}^{\ast }-\Omega _{c}^{\ast
}\sigma _{32}+\Omega _{c}\sigma _{32}^{\ast }=0,  \label{1c} \\
& &\left( i\frac{\partial }{\partial t}+d_{21}\right) \sigma _{21}-\Omega
_{p}\sigma _{32}^{\ast }+\Omega _{c}^{\ast }\sigma _{31}=0,  \label{1d} \\
& &\left( i\frac{\partial }{\partial t}+d_{31}\right) \sigma _{31}-\Omega
_{p}(\sigma _{33}-\sigma _{11})+\Omega _{c}\sigma _{21}=0,  \label{1e} \\
& &\left( i\frac{\partial }{\partial t}+d_{32}\right) \sigma
_{32}-\Omega _{c}(\sigma _{33}-\sigma _{22})+\Omega _{p}\sigma
_{21}^{\ast }=0,
\end{eqnarray}
\end{subequations}%
where $\Omega _{p}=\mathbf{e}_{p}\cdot
\mathbf{p}_{31}\mathcal{E}_{p}/\hbar $ and $\Omega
_{c}=\mathbf{e}_{c}\cdot \mathbf{p}_{32}\mathcal{E}_{c}/\hbar$ are
respectively the half Rabi frequencies of the probe and the
control fields, with $\mathbf{p}_{jl}$ being the electric dipole
matrix element associated with the transition from $|l\rangle$ to
$|j\rangle$. In Eq.~(\ref{BLO}), we have defined
$d_{21}=\Delta_{2}+i\gamma_{21}$,
$d_{31}=\Delta_{3}+i\gamma_{31}$, and
$d_{32}=(\Delta_{3}-\Delta_{2})+i\gamma_{32}$, where $\Delta_{2}$
and $\Delta_{3}$ are two- and one-photon detunings given respectively by
$\Delta_{2}=\delta_{2}+\mu_{21}f(t)Bx-\frac{1}{4}\alpha_{21} p^2
J_l^2(\sqrt{2b}r)$ and
$\Delta_{3}=\delta_{3}+\mu_{31}f(t)Bx-\frac{1}{4}\alpha_{31} p^2
J_l^2(\sqrt{2b}r)$, with
$\mu_{jl}=\mu_B(g_F^{j}m_F^{j}-g_F^{l}m_F^{l})/\hbar$ and
$\alpha_{jl}=(\alpha_{j}-\alpha_{l})/\hbar$. Here
$\delta_{2}=\omega_{p}-\omega_{c}-\omega_{21}$ and
$\delta_{3}=\omega_{p}-\omega_{31}$, where
$\omega_{jl}=(E_j-E_l)/\hbar$ with $E_j$ being the eigenenergy of
the state $|j\rangle$.
$\gamma_{jl}=(\Gamma_j+\Gamma_l)/2+\gamma_{jl}^{\rm col}$, where
$\Gamma_j=\sum_{j<l}\Gamma_{jl}$ with $\Gamma _{jl}$ being the
spontaneous emission decay rate from $|l\rangle$ to $|j\rangle$
and $\gamma _{jl}^{\rm col}$ being the dephasing rate reflecting
the loss of phase coherence between $|j\rangle$ and $|l\rangle$
without changing of population, as might occur by elastic
collisions.

The equation of motion for $\Omega _{p}(x,y,z,t)$ can be obtained
by the Maxwell equation, which under slowly-varying envelope
approximation reads
\begin{equation}\label{MAX}
i\left( \frac{\partial }{\partial z}+\frac{1}{c}\frac{\partial }{\partial t}%
\right) \Omega _{p}+\frac{c}{2\omega _{p}}\left(\frac{\partial
^{2}}{\partial x^{2}}+\frac{\partial ^{2}}{\partial
x^{2}}\right)\Omega _{p}+\kappa _{13}\sigma _{31}=0,
\end{equation}
where $\kappa _{13}=N_{a}\omega
_{p}|\mathbf{p}_{13}|^{2}/(2\varepsilon_{0}c\hbar )$ with
$N_{a}$ being the atomic concentration.

Our model can be easily realized by experiment. One of candidates
is the cold $^{85}$Rb atomic gas with the energy-levels in Fig.~\ref{fig1}(a)
assigned as $|1\rangle=|5^2S_{1/2},F=2\rangle$,
$|2\rangle=|5^2S_{1/2},F=3\rangle$, and
$|3\rangle=|5^2P_{1/2},F=2\rangle$. Then, the decay rates in the Bloch
Eq.~(\ref{BLO}) are given by
$\Gamma_{13}\approx\Gamma_{23}=5.9$ MHz and $\gamma_{13}^{\rm col}\approx\gamma_{23}^{\rm
col}=50$ Hz. In addition, we take $N_a=10^{12}$ cm$^{-3}$, then
$\kappa_{13}$ in the Maxwell Eq.~(\ref{MAX}) takes the value $1.0\times
10^{9}$ cm$^{-1}$s$^{-1}$. We shall use these system parameters
in the following calculations.


\section{(3+1)-dimensinal nonlinear envelope equation}

The base state solution (i.e. the steady-state solution for vanishing
$\Omega_p$) of the MB Eqs.~(\ref{BLO}) and (\ref{MAX})
is $\sigma_{11}=1$ and other $\sigma_{jl}$ are zero.
When a weak probe field (i.e. $\Omega_p$ is very small) is applied,
the system undergoes a linear evolution.
In this case, the MB Eqs.~(\ref{BLO}) and (\ref{MAX}) can be linearized
with the solution given by
\begin{subequations}
\label{linear}
\begin{eqnarray}
&&\Omega_{p}=A e^{i[K(\omega)z-\omega t]}, \\
&&\sigma_{j1}=\frac{\delta_{j3}(\omega+\delta_{2}+i\gamma_{21})
-\delta_{j2}\Omega_{c}^{\ast}}{D(\omega)}A e^{i(K(\omega)z-\omega t)},\quad (j=2,3)
\end{eqnarray}
\end{subequations}%
together with $\sigma_{jj}=\delta_{j1}$ and $\sigma_{32}=0$. Here
$A$ is a constant, $\delta_{jk}$ is Kronecker delta symbol, and
$K(\omega)$ is the linear dispersion relation of the system
\begin{equation}\label{Disp}
K(\omega)=\frac{\omega}{c}+\kappa_{13}\frac{\omega+\delta_{2}+i\gamma_{21}}{D(\omega)}
\end{equation}%
with
$D(\omega)=|\Omega_{c}|^{2}-(\omega+\delta_{2}+i\gamma_{21})(\omega+\delta_{3}+i\gamma_{31})$.
Here $\omega$ and $K(\omega)$ are respectively the deviations of
the frequency and wavenumber of the probe field~\cite{note1}.
In obtaining Eq.~(\ref{Disp}) we have neglected the transverse
diffraction effect which is usually negligible in the leading
order approximation of Eq.~(\ref{MAX}). For illustration, in the panels (a)
%
\begin{figure}[tbph]
\centering
\includegraphics[scale=0.75]{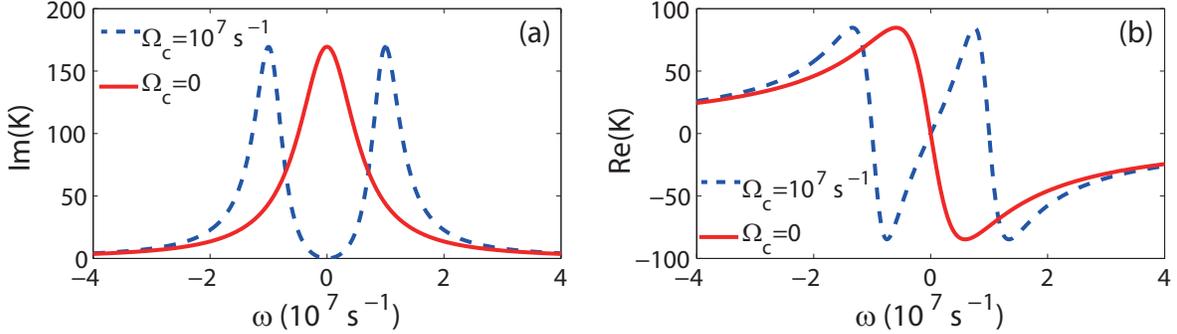}
\caption{{\protect\footnotesize (Color online) Im$(K)$ (a)
and Re$(K)$ (b) as functions of $\omega$. The dashed and
solid lines in each panel correspond to the presence
($\Omega_{c}=1.0\times 10^{7}$ ${\rm s}^{-1}$) and the absence
($\Omega_{c}=0$) of the control field, respectively. }}
\label{fig2}
\end{figure}
and (b) of Fig.~\ref{fig2} we have plotted the real part Re$(K)$
and the imaginary part Im$(K)$ as a function of $\omega$ for
$\Delta_{2}=\Delta_{3}=0$. The dashed and solid lines in the figure
correspond respectively to the absence ($\Omega_{c}=0$) and the presence
($\Omega_{c}=1.0\times 10^{7}$ ${\rm s}^{-1}$)
of the control field. One sees that when $\Omega_{c}=0$,
the probe field has a large absorption (the solid line of
panel (a)\,); however, when $\Omega_{c}\neq 0$ and increases
to a large value, a transparency window is opened in the probe-field
absorption spectrum (the dashed line of panel  (a)\,), and hence
the probe field can propagate in the resonant atomic system
with negligible absorption, a basic character of EIT. On the other hand,
for the large control field  the slope of Re$(K)$ is drastically
changed and steepened (see the dashed line of panel (b)\,)  which
results in a significant reduction of the group velocity of
the probe field (and hence slow light). All these interesting characters are
due to the quantum interference effect induced by the control
field~\cite{fle}.


However, although the absorption is largely suppressed by the EIT effect,
the probe pulse may still suffers a serious distortion during propagation
because of the existence of the dispersion and diffraction.
To avoid such distortion and obtain a long-distance propagation of
shape-preserving probe pulses, a natural idea is to use nonlinear effect to
balance the dispersion and diffraction. One of important shape-preserving
(3+1)-dimensional probe pulses is optical bullet.

For this aim, we first derive a (3+1)-dimensional nonlinear envelope equation
that includes the dispersion, diffraction, and nonlinearity of the system.
We take the asymptotic expansion
\begin{subequations}
\label{EXP}
\begin{eqnarray}
&&\sigma _{jk}=\sigma _{jk}^{(0)}+\epsilon
\sigma_{jk}^{(1)}+\epsilon^{2}\sigma_{jk}^{(2)}+\epsilon^{3}\sigma_{jk}^{(3)}+\cdots
,\quad (j,\,k=1,\,2,\,3),  \label{3a} \\
&&\Omega _{p}=\epsilon \Omega_{p}^{(1)}
+\epsilon ^{2}\Omega_{p}^{(2)}+\epsilon^{3}\Omega_{p}^{(3)}+\cdots, \\
&&d_{j1}=d_{j1}^{(0)}+\epsilon d_{j1}^{(1)}
+\epsilon^{2}d_{j1}^{(2)}+\cdots,\quad (j=2,\,3),\\
&&d_{32}=d_{32}^{(0)}+\epsilon
d_{32}^{(1)}+\epsilon^{2}d_{32}^{(2)}+\cdots,
\end{eqnarray}
\end{subequations}%
with $\sigma _{jk}^{(0)}=\delta_{j1}\delta_{k1}$.  Additionally,
we assume both the gradient magnetic field (\ref{SGfield}) and
the far-detuned laser field (\ref{Efield}) are of order of
$\epsilon$. Thus we have $d_{21}^{(0)}=\delta_{2}+i\gamma_{21}$,
$d_{31}^{(0)}=\delta_{3}+i\gamma_{31}$,
$d_{32}^{(0)}=\delta_{3}-\delta_{2}+i\gamma_{32}$,
$d_{21}^{(1)}=d_{31}^{(1)}=d_{32}^{(1)}=0$,
$d_{21}^{(2)}=\mu_{21}f(t_2)Bx_1-\frac{1}{4}\alpha_{21} p^2
J_l^2(\sqrt{2b}r_1)$,
$d_{31}^{(2)}=\mu_{31}f(t_2)Bx_1-\frac{1}{4}\alpha_{31} p^2
J_l^2(\sqrt{2b}r_1)$, and
$d_{32}^{(2)}=\mu_{32}f(t_2)Bx_1-\frac{1}{4}\alpha_{32} p^2
J_l^2(\sqrt{2b}r_1)$. Here $\epsilon $ is a dimensionless small
parameter characterizing the amplitude of the probe field. All
quantities on the right hand side of the expansion (\ref{EXP}) are
considered as functions of the multi-scale variables
$z_{l}=\epsilon^{m}z$ ($m=0$ to $2$), $t_{2}=\epsilon^{2}t$,
$x_{1}=\epsilon x$, $y_{1}=\epsilon y$, and
$r_1=\sqrt{x_1^2+y_1^2}$.

Substituting the expansions (\ref{EXP}) into the MB Eqs. (\ref{BLO}) and
(\ref{MAX}), and comparing the coefficients of
$\epsilon^m$ ($m=1,2,3\cdots$), we obtain a set of linear
but inhomogeneous equations which can be
solved order by order (see Appendix A for more details).

At the leading order ($m=1$), we have the solution in the linear
regime the same as that given by Eq.~(\ref{linear}).
However,  now $A$ is a yet to be determined envelope function
of the slow variables  $t_{2}$, $x_{1}$, $y_{1}$, and
$z_{j}$ ($j=1$, $2$).

At the next order ($m=2$), a divergence-free condition requires
$\partial A/\partial z_1=0$, i.e. $A$ is independent on $z_1$.
The second-order solution reads $\sigma _{31}^{(2)}=\sigma _{21}^{(2)}=0$,
$\sigma_{jj}^{(2)}=a_{jj}^{(2)}|A|^{2}e^{-\bar{\alpha}z_{2}}$
$(j=1,\,2)$, and
$\sigma_{32}^{(2)}=a_{32}^{(2)}|A|^{2}e^{-\bar{\alpha}z_{2}}$,
where
\begin{subequations}
\begin{eqnarray}
&&a_{11}^{(2)}=\frac{\left[ i\Gamma _{23}-2|\Omega _{c}|^{2}\left( \dfrac{1}{%
d_{32}^{(0)}}-\dfrac{1}{d_{32}^{(0)\ast }}\right) \right]G
-i\Gamma _{13}|\Omega _{c}|^{2}\left( \dfrac{1}{Dd_{32}^{(0)\ast }}-\dfrac{1%
}{D^{\ast }d_{32}^{(0)}}\right) }{i\Gamma _{13}|\Omega _{c}|^{2}\left(
\dfrac{1}{d_{32}^{{(0)}\ast }}-\dfrac{1}{d_{32}^{(0)}}\right) },  \nonumber\\
&&a_{22}^{(2)}=\frac{G-i\Gamma
_{13}a_{11}^{(2)}}{i\Gamma _{13}}, \nonumber \\
&&a_{32}^{(2)}=\frac{1}{d_{32}^{(0)}}\left[ \frac{\Omega _{c}}{D^{\ast }}%
-\Omega _{c}(a_{11}^{(2)}+2a_{22}^{(2)})\right] ,  \nonumber
\end{eqnarray}
\end{subequations}
and
$\bar{\alpha}=\epsilon^{-2}\alpha=\epsilon^{-2}\text{Im}[K(\omega
)]$, with
$G=(\omega+d_{21}^{(0)\ast})/D^{\ast}-(\omega+d_{21}^{(0)})/D$ .

With the above results we proceed to the third order ($m=3$). The
divergence-free condition in this order yields the nonlinear
equation for the envelope function $A$:
\begin{equation}
i\left(\frac{\partial }{\partial
z_{2}}+\frac{1}{V_g}\frac{\partial }{\partial
t_{2}}\right)A+\frac{c}{2\omega _{p}}\left(
\frac{\partial ^{2}}{\partial x_{1}^{2}}+\frac{\partial ^{2}}{\partial
y_{1}^{2}}\right)A+ U(t_2,r_1)A-W|A|^{2}Ae^{-2\bar{\alpha}
z_{2}}=0.  \label{ORD3}
\end{equation}
where $V_{g}=(\partial K/\partial \omega)^{-1}$  is the group velocity of $A$,
and
\begin{eqnarray}
W
&=&-\kappa_{13}\frac{\Omega_{c}a_{32}^{(2)\ast}+\left(\omega+d_{21}^{(0)}\right)
(2a_{11}^{(2)}+a_{22}^{(2)})}{D(\omega)}, \nonumber\\
U
&=&\kappa_{13}\frac{d_{21}^{(2)}|\Omega_{c}|^{2}
+d_{31}^{(2)}(\omega+d_{21}^{(0)})^{2}}{D^{2}(\omega)}. \nonumber
\end{eqnarray}
Here $W$ is proportional to Kerr coefficient characterizing
self-phase modulation (SPM) effect and
$U$ represents a trapping potential with the form
\be \label{potential} U(t_2,r_1)={\cal B} f(t_2)Bx_1+{\cal E}
p^2 J_l^2\left(\sqrt{2\bar{b}}r_1\right),
\ee
where ${\cal
B}=\kappa_{13}[|\Omega_{c}|^2\mu_{21}+(\omega+d_{21}^{(0)})^2\mu_{31}]/D^{2}(\omega)$
and ${\cal
E}=-\kappa_{13}[|\Omega_{c}|^2\alpha_{21}+(\omega+d_{21}^{(0)})^2\alpha_{31}]/[4D^{2}(\omega)]$,
with $\bar{b}=\epsilon^{-2}b$. We see that the potential $U$ consists of two
parts, contributed respectively by the time-dependent
gradient magnetic field (\ref{SGfield}) and the
far-detuned laser field (\ref{Efield}).

When returning to original variables, Eq.~(\ref{ORD3})
can be written into the dimensionless form
\begin{equation}\label{NLS1}
i\left[\left(\frac{\partial }{\partial
s}+\frac{1}{g}\frac{\partial }{\partial \tau}\right)
+d_0\right]{\cal A}+\frac{1}{2}\left(\frac{\partial^{2} }{\partial
\xi^{2}}+\frac{\partial^{2} }{\partial \eta^{2}}\right){\cal A}
+{\cal U}(\tau,\rho){\cal A} +d_1|{\cal A}|^{2}{\cal A}=0,
\end{equation}
where we have introduced the dimensional variables
$s=z/L_{\rm Diff}$, $\tau=t/\tau_0$,
$(\xi,\eta)=(x,y)/R_{\bot}$, $g={\rm Re}(V_g)\tau_0/L_{\rm Diff}$,
$\rho=\sqrt{\xi^2+\eta^2}$, ${\cal A}=(\Omega_{p}/U_0) e^{-i{\rm
Re}[K|_{\omega=0}]z}$, and ${\cal U}(\tau,\rho)=UL_{\rm Diff}$,
with $L_{\rm Diff}=\omega_p R_{\bot}^2/c$ being the diffraction
length. We have also introduced the characteristic absorption
and nonlinearity lengths respectively defined by $L_{\rm
Abs}=1/\alpha|_{\omega=0}$
and $L_{\rm Nonl}=1/(|W|U_0^2)$, with $U_0$ being
the typical Rabi frequency. Dimensionless coefficients
$d_j$ ($j=0,1$) in Eq.~(\ref{NLS1}) are given by  $d_0=L_{\rm Diff}/L_{\rm Abs}$
and $d_{1}=-{\rm sign}({\rm Re}(W))L_{\rm Diff}/L_{\rm Nonl}$, respectively.
Notice that when deriving Eq.~(\ref{NLS1}) we have set $\omega=0$ and assumed the
imaginary part of $V_g$ and $W$ can be made much smaller than their
corresponding real part, which can be indeed achieved
because of the EIT effect in the system (see a typical example given
in the next section).

Equation~(\ref{NLS1}) has the form of (3+1)-dimensional
~\cite{note2} nonlinear Schr\"{o}dinger
(NLS) equation. However, it is still too complicated for
analytical and numerical studies. For simplicity, we neglect the small absorption
(i.e disregarding the term proportional to $d_0$). Furthermore, we assume Re($W)<0$ and
$L_{\rm Diff}=L_{\rm Nonl}$ (i.e. $d_1=1$; see the a typical example given
in the next section). Then Eq. (\ref{NLS1}) can be written into
\be\label{NLS2}
i\left(\frac{\partial }{\partial s}+\frac{1}{g}\frac{\partial
}{\partial \tau}\right){\cal
A}+\frac{1}{2}\left(\frac{\partial^{2} }{\partial
\xi^{2}}+\frac{\partial^{2}}{\partial \eta^{2}}\right){\cal
A}+[F(\tau)\xi+PJ_l^2(\sqrt{2\beta}\rho)]{\cal A}+|{\cal
A}|^{2}{\cal A}=0,
\ee
where $F(\tau)={\cal B} f(\tau)BR_{\bot}L_{\rm Diff}$, $P={\cal E}
p^2 L_{\rm Diff}$, and $\beta= bR_{\bot}^2$.


\section{Ultraslow helical optical solitons and their stability}

We now consider the evolution of a probe wave packet having the form
\be\label{ansatz}
{\cal A}(\tau,\xi,\eta,s)=\psi(\tau,s)\varphi(\tau,\xi,\eta), \ee
with
\be\label{gau}
\psi(\tau,s)=\frac{1}{\sqrt[4]{2\pi\sigma^2}}e^{-(s-g\tau)^2/(4\sigma^2)}
=\frac{1}{\sqrt[4]{2\pi\sigma^2}}e^{-(z-{\rm Re}(V_g)
t)^2/(4\sigma^2L_{\rm Diff}^2)}, \ee
where $\sigma$ is a free real parameter. Obviously,  $\psi(\tau,s)$
describes a shape-preserving Gaussian pulse propagating with the
group velocity Re($V_g$) along the $z$-axis.
Substituting (\ref{ansatz}) into Eq.
(\ref{NLS2}) and integrating over the variable $s$, we obtain the equation
for $\varphi(\tau,\xi,\eta)$:
\bea\label{NLS3}
&& i\frac{1}{g}\frac{\partial}{\partial
\tau}\varphi+\frac{1}{2}\left(\frac{\partial^{2} }{\partial
\xi^{2}}+\frac{\partial^{2} }{\partial
\eta^{2}}\right)\varphi+[F(\tau)\xi+PJ_l^2(\sqrt{2\beta}\rho)]
\varphi+\frac{1}{2\sqrt{\pi}\sigma}|\varphi|^{2}\varphi=0.
\eea
If the solutions for $\varphi(\tau,\xi,\eta)$ localized
in both $\xi$ and $\eta$ directions can be found, the solutions for
probe-field envelope ${\cal A}$ will be optical bullets
localized in all three spatial dimensions.

For the convenience of the following calculations, we take a
set of realistic system parameters given by $\Omega_c=1.6\times10^7$ s$^{-1}$,
$\delta_2=-8.0\times10^5$ s$^{-1}$, $\delta_3=4.0\times10^7$
s$^{-1}$, $R_{\bot}=5.0\times10^{-3}$ cm, $\tau_0=6.0$ $\mu$s, and
$U_0=5.6\times10^6$ s$^{-1}$.  We then have
$K|_{\omega=0}=-2.78+i0.05$ cm$^{-1}$,
$(\partial K/\partial\omega)|_{\omega=0}=(3.09-i0.10)\times10^{-6}$
cm$^{-1}$ s,
and $W=(-1.62+i0.04)\times10^{-14}$ cm$^{-1}$ s$^2$. Note that the
imaginary parts of these quantities are indeed much smaller
than their corresponding real parts, as we indicated above.
The characteristic lengths of the system are
$L_{\rm Abs}=20.58$ cm, $L_{\rm Diff}=1.98$ cm,
and $L_{\rm Nonl}=1.98$ cm, leading to $d_0=0.09$
and $d_1=1$. The group velocity reads
\begin{equation}
{\rm Re}(V_g)=1.08\times10^{-5}c,
\end{equation}
which is much slower than the light speed in the vacuum and gives $g\approx 1$
in Eq.~(\ref{NLS3}).

\begin{figure}[tbph]
\centering
\includegraphics[scale=0.8]{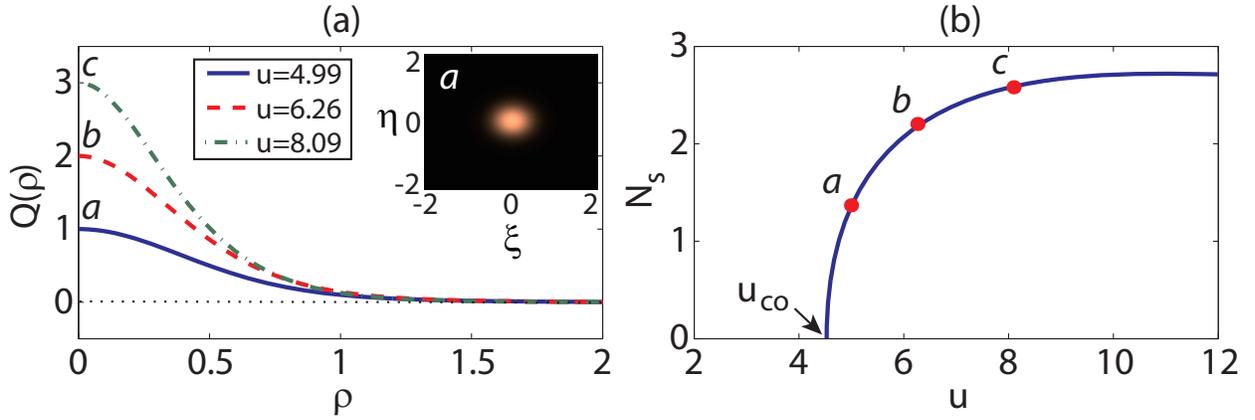}
\caption{{\protect\footnotesize (Color online) (a): Profile $Q$ of
the lowest-order ($l=0$) optical bullet for $u=4.99$ (solid line,
labeled by ``a''), 6.26 (dashed line, labeled by ``b''), and
8.09 (dot-dashed line, labeled by ``c''). Inset: Intensity
distribution of the optical bullet for $u=4.99$
in the $(\xi,\eta)$ plane (corresponding to
the profile ``a'').
(b): The norm of the optical bullet $N_s$ as a function of $u$. The cutoff
value of $u$ is $u_{\rm co}\approx 4.529$. The solid circles indicated by
``a'', ``b'', and ``c'' are relevant to the profiles ``a'', ``b'', and ``c''
in the panel (a). }}
\label{fig3}
\end{figure}

We now turn to seek nonlinear localized solutions of Eq.~(\ref{NLS3}). We first consider
the situation in the absence of the gradient magnetic field (i.e $F(\tau)=0$). The case
of the presence of the gradient magnetic field (i.e $F(\tau)\neq 0$) will be considered
in the next section. Assuming $\varphi=Q(\rho) \exp(-iu\tau)$, Eq.~(\ref{NLS3})
reduces to
\be \label{NLS4}
\frac{d^2 Q}{d\rho^2}+\frac{1}{\rho}\frac{d Q}{d\rho}
+2\left[PJ_l^2(\sqrt{2\beta}\rho)-u\right] Q+2Q^3=0,
\ee
which can be solved by a shooting method. The lowest-order optical
bullet solution whose intensity maximum coincides with the
center of the zero-order (i.e. $l=0$) Bessel optical lattice potential
(the corresponding light intensity has been illustrated in the
left part of Fig.~\ref{fig1}(c)\,). Shown in Fig.~\ref{fig3}
%
%
is the result of numerical simulation for the lowest-order ($l=0$)
optical bullet and its stability. Fig.~\ref{fig3}(a) gives profile
$Q$ of the
optical bullets for different values of $u$. In the simulation,
$P=10$ and $\beta=2$ in Eq.~(\ref{NLS4}) for $l=0$ have been chosen. The
solid, dashed, and dotted-dashed lines in Fig.~\ref{fig3}(a) are
for $u=4.99$, 6.26, and 8.09, respectively. The inset shows the
intensity distribution of the optical bullet for $u=4.99$ in the
($\xi,\eta$) plane. We see that the optical-bullet amplitude grows
when $u$ increases.

The norm of the optical bullet, defined by
$N_{s}=2\pi\int_0^{\infty }Q^2(\rho) \rho d\rho$,  is found to be a
monotonically growing function of $u$, i.e. $\partial N_s/\partial u>0$
(see Fig.~\ref{fig3}(b)\,), which implies the optical
bullet is stable according to Vakhitov-Kolokolov
criterion~\cite{VK}. However, the optical
bullet exists only for $u\ge u_{\rm co}$.
The cutoff value $u_{\rm co}$ grows when the strength $P$ of the
Bessel optical lattice potential increases, and the optical
bullet's norm vanishes for $u \rightarrow u_{\rm co}$.
For $P=10$, we obtain $u_{\rm co}\approx 4.529$.
The solid circles indicated by ``a'', ``b'', and ``c'' in the
panel (b) are  relevant to the profiles ``a'', ``b'', and ``c''
in the panel (a).

\begin{figure}[tbph]
\centering
\includegraphics[scale=0.75]{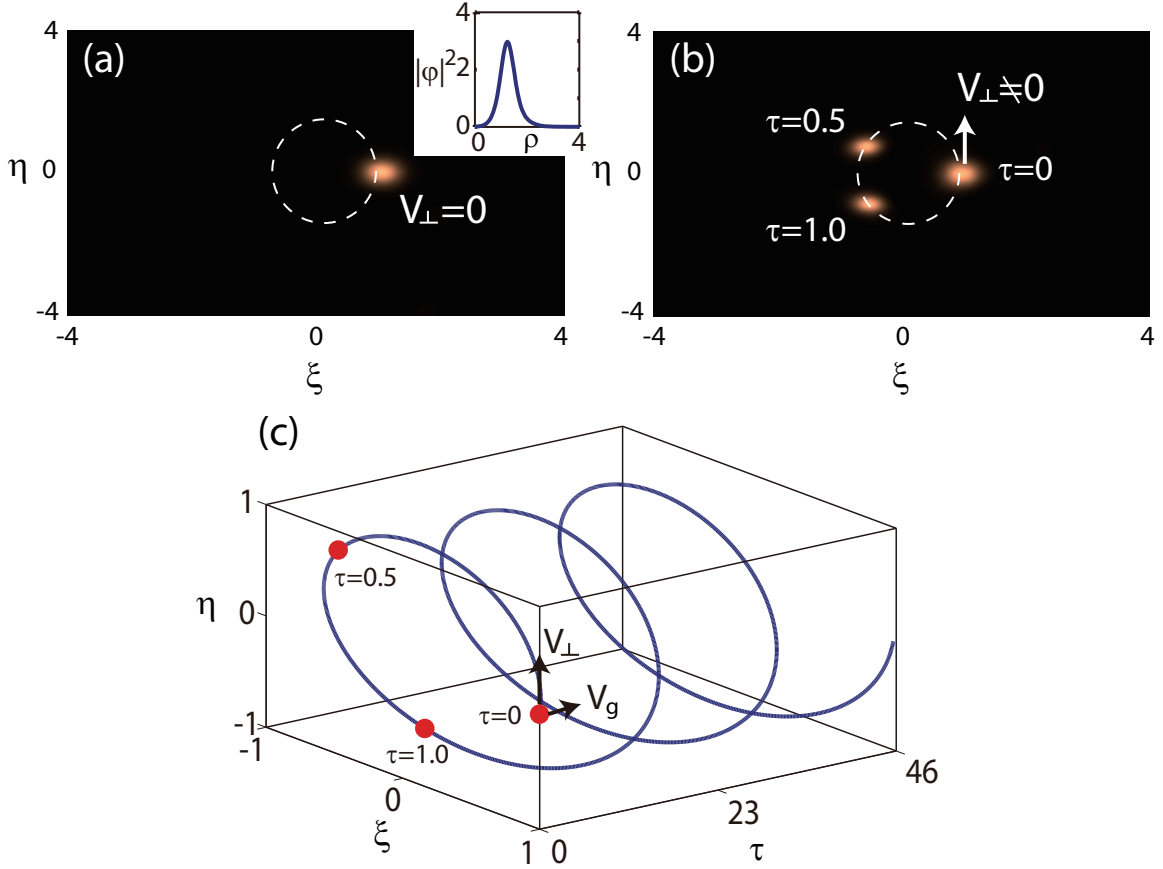}
\caption{{\protect\footnotesize (Color online) (a): Intensity
distribution in ($\xi,\eta$) plane of a stationary first-order
($l=1$) optical bullet without transverse velocity (i.e.
$V_{\perp}=0$). The radius of the first ring (denoted by dashed
circle) of the first-order Bessel lattice is $R_L\approx 0.92$.
The inset shows the distribution $|\varphi|^2$ as a function of
$\rho$. (b): Intensity distribution of a rotational first-order
optical bullet with initial transverse velocity
$V_{\perp}=1.07\times10^{-7}$ $c$ for $\tau=0$, $0.5$, and $1.0$,
respectively. (c): The helical motional trajectory (solid line) of
the ultraslow optical bullet, which has the longitudinal velocity
$V_{g}$ and the transverse velocity $V_{\perp}$. The solid circles
indicate the position of the the optical bullet for $\tau=0$,
$0.5$, and $1.0$, respectively. }} \label{fig4}
\end{figure}

We have also found high-order (i.e. $l\ge 1$) optical bullets in the
system, which are stable nonlinear localized solutions in the presence
of the  high-order ($l\ge 1$) Bessel optical lattice. Such high-order optical bullets
are trapped in the rings of off-center radial maxima of the Bessel
optical lattice, where the refractive index contributed by the
optical lattice are maximum. An example of a 1-order optical
bullet in the 1-order Bessel optical lattice (the corresponding light
intensity has been illustrated in the right part of
Fig.~\ref{fig1}(c)\,) has been shown in Fig.~\ref{fig4}(a).
%
%
The optical bullet locates in the first ring of the Bessel
optical lattice, and has no transverse (i.e. $\xi$- and $\eta$-direction)
velocity (i.e. $V_{\perp}=0$). The stability of the optical bullet
has been verified using direct numerical simulations by
considering the time evolution of the optical bullet with added random
perturbations of relative amplitude up to 10\% level.

The optical bullets obtained above have no transverse velocity,
but, from expressions (\ref{ansatz}) and (\ref{gau}), they have a
longitudinal velocity $V_g$ along the $z$-direction. We now show
that, in fact, the high-order (i.e. $l\ge 1$) optical bullets can
acquire a transverse velocity, and hence they can display a
helical motion in the 3-dimensional space. This is possible
because in each ring of the Bessel optical lattice the potential
energy is degenerate, therefore an optical bullet will move around
the ring with the minimum potential energy if an initial
transverse velocity $V_{\perp}$ tangent to the ring is given
\cite{rotary}. Shown in Fig.~\ref{fig4}(b) is the result of the
rotation of a 1-order optical bullet with $V_{\perp}=1.07 \times
10^{-7}$ $c$ for $\tau=0$, 0.5, and 1 (corresponding to $t$=0, 3,
and 6 $\mu$s), respectively. Since now the optical bullet has the
longitudinal velocity $V_g$ and also the transverse velocity
$V_{\perp}$, it makes a helical motion in the 3-dimensional space.
Fig.~\ref{fig4}(c) shows such helical motion of the optical
bullet, where the red solid circles indicate the position of the
the optical bullet for $\tau=0$, $0.5$, and $1.0$, respectively.
Because both $V_g$ and $V_{\perp}$  are much smaller than $c$, the
nonlinear localized structure obtained here is indeed an {\it ultraslow
helical optical bullet}.

\section{Acceleration of ultraslow helical optical bullets}

Since the SG gradient magnetic fields can be used to
change the propagating direction of optical
beams~\cite{KW}, it can also be used to change the propagating velocity of optical
pulses. We now study the acceleration of the ultraslow helical optical bullets.
To this end, both the Bessel optical lattice and the time-dependent SG
magnetic field must be applied simultaneously.

In order to describe the optical-bullet acceleration analytically, we
assume that the ring-shaped trap formed by the Bessel optical lattice
is narrow enough so that it ensures a quasi one-dimensional distribution
of the light intensity of the optical bullet along the
ring-shaped trap. On the other hand, the ring-shaped trap is also deep enough so
that we can look for the solution with the form
\be \varphi(\tau,\rho,\theta)=\Phi(\rho)\Xi(\tau,\theta), \ee
where $\Phi$ is the normalized ground state of the eigenvalue
problem
\be
\frac{1}{2\rho}\frac{d}{d\rho}\rho\frac{d}{d\rho}\Phi(\rho)
+PJ_l^2(\sqrt{2\beta}\rho)\Phi(\rho)=E_r\Phi(\rho),
\ee
with $E_r$  the eigenvalue and $\Phi$ satisfying
the normalization condition $\int_0^{\infty}\Phi^2\rho
d\rho=1/(2\pi)$.  Notice that the eigenvalue,
and also $\Phi(\rho)$, depend on $l$. Here for
simplicity we focus on the special situation $l=1$.
After integrating over the variable $\rho$ and writing the
equation into polar coordinates $(\rho,\,\theta)$, Eq.
(\ref{NLS3}) becomes
\be\label{NLS5}
i\frac{\partial}{\partial
\tau}\Xi+\frac{R_1}{2}\frac{\partial^{2} }{\partial
\theta^{2}}\Xi+[E_r+R_2F(\tau)\cos\theta]\Xi
+\frac{R_3}{2\sqrt{\pi}\sigma}|\Xi|^{2}\Xi=0,
\ee
where $R_j$ ($j=1,2,3$) are given by
$R_1=\int_0^{\infty}(\Phi^2/\rho) d\rho$,
$R_2=\int_0^{\infty}\Phi^2\rho^2 d\rho$, and
$R_3=\int_0^{\infty}\Phi^4\rho^2 d\rho$, respectively.

Let $\tilde{\Xi}=\sqrt{\frac{R_3}{2\sqrt{\pi}\sigma}}e^{-iE_r\tau}\Xi$,
Eq.~(\ref{NLS5}) can be further simplified to
\be\label{NLS6}
i\frac{\partial}{\partial
\tau}\tilde{\Xi}+\frac{1}{2}\frac{\partial^{2} }{\partial
\vartheta^{2}}\tilde{\Xi}+\tilde{F}(\tau)\cos(\sqrt{R_1}
\vartheta)\tilde{\Xi}+|\tilde{\Xi}|^{2}\tilde{\Xi}=0,
\ee
where $\vartheta=\theta/\sqrt{R_1}$ and $\tilde{F}=R_2F$.
If the time-dependent gradient magnetic field is absent, i.e.
$\tilde{F}(\tau)=0$, we can obtain the exact solution of Eq.~(\ref{NLS6})
expressed by a Jacobi elliptic function, which, when taking the modulus
of the Jacobi elliptic function as unity,  reduces to a bright
soliton moving with a constant velocity $v$:
\be\label{sol}
\tilde{\Xi}(\tau,\vartheta)=e^{iv(\vartheta-v\tau)-i\Omega
\tau}\eta\,{\rm sech}[\eta(\vartheta-v\tau)],
\ee
with $\Omega=-\eta^2/2+v^2/2$. We should bear in mind that $v$ in
Eq. (\ref{sol}) is an angular velocity which is related to the
transverse velocity by the relation
 $V_{\perp}=(v\sqrt{R_1})R_L$ with $R_L$
being the radius of the first (i.e. $l=1$)
ring of the Bessel lattice. Returning to original
variables, the soliton solution (\ref{sol}) reads
\be
\Xi(\tau,\theta)=\sqrt{\frac{2\sqrt{\pi}\sigma}{R_3}}
e^{iE_r\tau+iv(\theta/\sqrt{R_1}-v\tau)-i\omega
\tau}\eta\,{\rm sech}\left[\eta\left(\frac{\theta}{\sqrt{R_1}}-v\tau\right)\right].
\ee

If the time-dependent SG gradient magnetic field is present, i.e.
$\tilde{F}(\tau)\neq0$, the velocity $v$ of the soliton is no longer
preserved. Based on the solution (\ref{sol}),
under adiabatic approximation the solution of the perturbed equation
(\ref{NLS5}) can be assumed as the form
\be\label{sol1}
\tilde{\Xi}(\tau,\vartheta)=e^{iv(\tau)[\vartheta-\Theta(\tau)]-i\omega(\tau)
\tau}\eta\,{\rm sech}(\eta[\vartheta-\Theta(\tau)]),
\ee
where $\Theta(\tau)$ is a time-dependent function yet to be determined.
Using the variational method employed in \cite{ST}, it is easy to obtain
the equation of motion for $\Theta(\tau)$:
\be\label{Th}
\frac{d^2\Theta}{d\tau^2}=-\sqrt{R_1}\tilde{F}(\tau)\sin(\sqrt{R_1}\Theta).
\ee

We are interested in the acceleration of the soliton  when it
undergoes a rotational motion along the ring. Actually, this
acceleration can be achieved by using a
steplike time dependence of $\tilde{F}(\tau)$ \cite{BK}. To this end we
assume that the soliton is initially centered at $\Theta(\tau=0)=0$
and require $\tilde{F}(\tau)$ to be zero for time intervals such
that $\Theta(\tau)\in [2q\pi/\sqrt{R_1},(2q+1)\pi/\sqrt{R_1}]$ and
to be a constant $\tilde{F}_0>0$ for $\Theta(\tau)\in
[(2q+1)\pi/\sqrt{R_1},(2q+2)\pi/\sqrt{R_1}]$ with
$q=0,1,2,\cdots$. In this way, the soliton acquires an acceleration
at each time because the force contributed by the SG gradient magnetic field
(see the right hand side of Eq.~(\ref{Th})\,) is always positive.

Now our task is to find time intervals $T_n=[\tau_n,\tau_{n+1}]$
($n=0,\,1,\,2,\cdots$), with $\tau_0=0$, such that
$\tilde{F}(\tau)=0$ for $\tau\in T_{2q}$ and
$\tilde{F}(\tau)=\tilde{F}_0$ for $\tau\in T_{2q+1}$. During the
time intervals $T_{2q}$, the soliton has to cross intervals
$[2q\pi/\sqrt{R_1},(2q+1)\pi/\sqrt{R_1}]$ with a constant velocity
while during the time intervals $T_{2q+1}$, the bullet has to
cross intervals $[(2q+1)\pi/\sqrt{R_1},(2q+2)\pi/\sqrt{R_1}]$ with
a growing velocity. By solving Eq. (\ref{Th}), we obtain
\be\label{T1}
T_{2q}=\tau_{2q+1}-\tau_{2q}=\frac{\pi}{\sqrt{R_1}v_{2q}},
\ee
where $v_{2q}$ is the velocity at the point $2q\pi/\sqrt{R_1}$,
and
\be\label{T2}
T_{2q+1}=\tau_{2q+2}-\tau_{2q+1}=\int_{\pi/\sqrt{R_1}}^{2\pi/\sqrt{R_1}}
\frac{d\Theta}{\sqrt{2\tilde{F}_0[\cos(\sqrt{R_1}\Theta)+1]+v_{2q+1}^2}},
\ee
where $v_{2q+1}$ is the velocity at the point
$(2q+1)\pi/\sqrt{R_1}$. In addition, we have $v_{2q+1}=v_{2q}$ and
$v_{2q+2}=\sqrt{4\tilde{F}_0+v_{2q+1}^2}$.

In a mechanical viewpoint, the acceleration of the soliton is caused by the
magnetic force exerted by the SG gradient magnetic field, and hence the soliton
possesses an effective magnetic moment~\cite{KW,HH}.
Note that the probe field is proportional to
${\cal A}$ (i.e. (\ref{ansatz})\,) which has a Gaussian factor $\psi$
(see (\ref{gau})\,), thus ${\cal A}$ is an optical bullet bounded in all
three spatial directions and displays an accelerated motion along the first ring
of the Bessel optical lattice. Since the optical bullet has also an
untraslow motional velocity $V_g$ in the longitudinal (i.e. $z$) direction,
its transverse acceleration can be observed for a very short medium
length. Similarly, one can also realize a deceleration of the optical
bullet along the ring by designing an appropriate force sequence, i.e.
another steplike time dependence of $\tilde{F}(\tau)$.

\begin{figure}[tbph]
\centering
\includegraphics[scale=0.75]{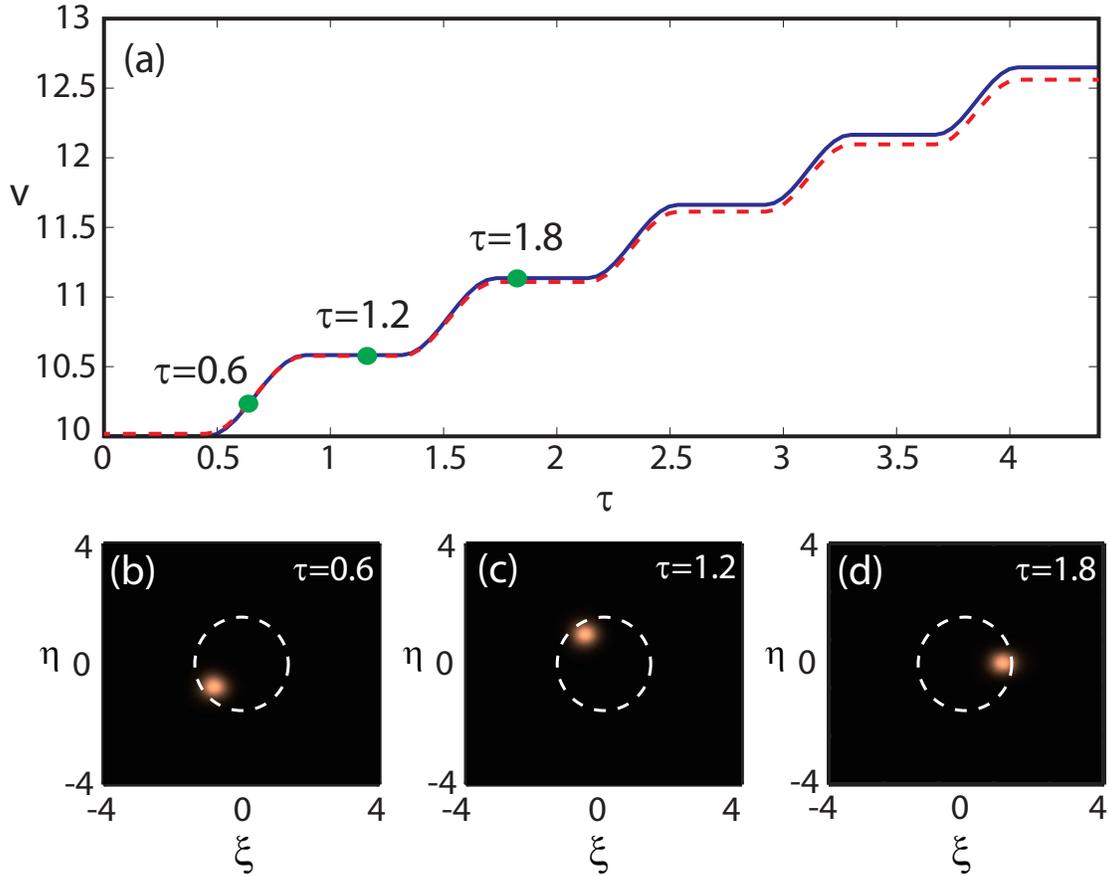}
\caption{{\protect\footnotesize (Color online) Transverse acceleration
of the optical bullet. (a): The angular velocity $v$ of the optical bullet
rotating along the first ring
of the first-order Bessel optical lattice as a function
of $\tau$. The solid (dashed) line is the result obtained from
Eq.~(\ref{Th}) (Eq.~(\ref{NLS5})\,) for $v_0=v(\tau=0)=10.0$
and $\tilde{F}_0=3.0$.  The large solid circles correspond to $\tau=0.6$,
1.2, and 1.8 , respectively.
(b), (c), (d): The light intensity distributions (positions) of the optical bullet
in the first ring of the Bessel lattice
(denoted by dashed circle with radius $R_L\approx0.92$)
for $\tau=0.6$, $1.2$, and $1.8$ in $(x,y)$ plane, respectively.
}}
\label{fig5}
\end{figure}

In Fig.~\ref{fig5}(a)
we compare the result of the solution obtained from Eq.~(\ref{Th})
with the result of the numerical simulation from Eq.~(\ref{NLS5}) for
the first-order optical bullet with $v_0=10.0$ and $\tilde{F}_0=3.0$.
During the simulation, $T_{2p}$ and $T_{2p+1}$ are
obtained by Eqs.~(\ref{T1}) and (\ref{T2}). One sees that
the solution obtained from
Eq.~(\ref{Th}) (solid line) and the numerical simulation (dashed line)
are matched quite well. The light intensity distributions (positions)
of the optical bullet in the first ring
of the Bessel lattice (denoted by dashed circle with radius $R_L\approx0.92$)
for $\tau=0.6$, 1.2, and 1.8
(corresponding to 3.6, 7.2, and 10.8 $\mu$s) are
respectively depicted in the panels (b), (c), and (d) of Fig.~\ref{fig5},
with the corresponding velocities  indicated by the large solid circles
in the panel (a). The transverse velocity of the
optical bullet can accelerate from $v_1=10.0$ (i.e. $V_\bot\approx1.8\times10^{-7}$ $c$)
to $v_2=10.6$ ($V_\bot\approx1.9\times10^{-7}$ $c$) in an atomic sample with the length
$L=V_gT_1\tau_0\approx0.85$ cm.

Using Poynting's vector, it is easy to calculate the input power needed
for generating the ultraslow helix optical bullets described above,
which is estimated as $P\approx0.2\times10^{-3}$ mW. Thus for
producing such optical bullets very low light intensity is required.
This is drastic contrast to conventional optical media such as
glass-based optical fibers, where ps or fs laser pulses are
usually needed to reach a very high peak power to bring out the
enough nonlinear effect needed for the formation of optical bullets~\cite{Ait}.


\section{Conclusion}

In this article, we have proposed a scheme to generate ultraslow
(3+1)-dimensional helical optical bullets in a
resonant three-level $\Lambda$-type atomic gas via EIT.
We show that due to EIT effect the helical optical
bullets can propagate with an ultraslow velocity up to
$10^{-5}$ $c$ in the longitudinal
direction and a slow rotational motion (with velocity
$10^{-7}$ $c$) in transverse directions. The generation
power of such optical bullets can be lowered to magnitude of
microwatt, and their stability can be achieved by using a Bessel optical
lattice formed by a far-detuned laser field. We have also demonstrated
that the transverse rotational motion of the optical
bullets can be accelerated by applying a time-dependent SG
magnetic field. Because of the untraslow velocity in the longitudinal direction,
a significant acceleration of the rotational motion of optical
bullets may be observed for a very short medium length.
Due to their interesting features, the ultraslow helical optical
bullets obtained here may become candidates for light
information processing and transmission at a very
weak light level.


\acknowledgments This work was supported by the NSF-China under
Grant Numbers 11174080 and 11105052, and by the Open Fund from the
State Key Laboratory of Precision Spectroscopy, ECNU.

\appendix


\section{The linear equations for each orders}

The MB Eqs.~(\ref{BLO}) and (\ref{MAX}) can be solved by standard method of
multiple-scales~\cite{huang}.
Substituting the expansion (\ref{EXP}) into the Eqs.~(\ref{BLO}) and
(\ref{MAX}) and comparing the coefficients of
$\epsilon^m$, we obtain the set of linear but inhomogeneous equations
\begin{subequations}
\label{App1}
\begin{eqnarray}
&&i\left( \frac{\partial }{\partial
z_{0}}+\frac{1}{c}\frac{\partial }{\partial t_{0}}\right)
\Omega_{p}^{(m)}+\kappa_{13}\sigma_{31}^{(m)}=M^{(m)},  \label{4a} \\
&&\left( i\frac{\partial }{\partial t_{0}}+d_{21}^{(0)}\right) \sigma_{21}^{(m)}+\Omega_{c}^{\ast }\sigma_{31}^{(m)}=N^{(m)},  \label{4b} \\
&&\left( i\frac{\partial }{\partial t_{0}}+d_{31}^{(0)}\right)
\sigma_{31}^{(m)}+\Omega_{p}^{(m)}+\Omega_{c}\sigma_{21}^{(m)}=P^{(m)},
\label{4c} \\
&&\left( i\frac{\partial }{\partial t_{0}}+d_{32}^{(0)}\right)
\sigma_{32}^{(m)}+\Omega_{c}(\sigma_{11}^{(m)}+2\sigma_{22}^{(m)})=Q^{(m)},
\label{4d} \\
&&i\frac{\partial }{\partial
t_{0}}\sigma_{11}^{(m)}+i\Gamma_{13}\sigma_{11}^{(m)}+i\Gamma_{13}\sigma_{22}^{(m)}=Y^{(m)},
\label{4e} \\
&&i\frac{\partial }{\partial
t_{0}}\sigma_{22}^{(m)}+i\Gamma_{23}\sigma_{22}^{(m)}+i\Gamma_{23}\sigma_{11}^{(m)}+\Omega_{c}^{\ast
}\sigma_{32}^{(m)}-\Omega_{c}\sigma_{32}^{\ast (m)}=Z^{(m)},
\label{4f}
\end{eqnarray}
\end{subequations}%
where the explicit expressions of $M^{(m)}$, $N^{(m)}$, $P^{(m)}$,
$Q^{(m)}$, $Y^{(m)}$ and $Z^{(m)}$ ($m=1,2,3,\cdots$) are given as
\begin{subequations}
\begin{eqnarray}
&&M^{(1)}=0,\,\,\,\,\,\,\,\,M^{(2)}=-i\left( \frac{\partial }{\partial z_{1}}%
+\frac{1}{c}\frac{\partial }{\partial t_{1}}\right) \Omega _{p}^{(1)}, \\
&&M^{(3)}=-i\left( \frac{\partial }{\partial z_{1}}+\frac{1}{c}\frac{%
\partial }{\partial t_{1}}\right) \Omega _{p}^{(2)}-i\frac{\partial }{%
\partial z_{2}}\Omega _{p}^{(1)}-\frac{c}{2\omega _{p}}\frac{\partial ^{2}}{%
\partial x_{1}^{2}}\Omega _{p}^{(1)}, \\
&&N^{(1)}=0,\,\,\,\,\,\,\,\,N^{(2)}=-i\frac{\partial }{\partial t_{1}}\sigma
_{21}^{(1)}+\Omega _{p}^{(1)}\sigma _{32}^{(1)\ast }, \\
&&N^{(3)}=-i\frac{\partial }{\partial t_{1}}\sigma _{21}^{(2)}+\Omega
_{p}^{(1)}\sigma _{32}^{(2)\ast }+\Omega _{p}^{(2)}\sigma _{32}^{(1)\ast
}-d_{21}^{(2)}\sigma _{21}^{(1)}, \\
&&P^{(1)}=0,\,\,\,\,\,\,\,\,P^{(2)}=-i\frac{\partial }{\partial t_{1}}\sigma
_{31}^{(1)}-\Omega _{p}^{(1)}(2\sigma _{11}^{(1)}+\sigma _{22}^{(1)}), \\
&&P^{(3)}=-i\frac{\partial }{\partial t_{1}}\sigma _{31}^{(2)}-\Omega
_{p}^{(1)}(2\sigma _{11}^{(2)}+\sigma _{22}^{(2)})-\Omega _{p}^{(2)}(2\sigma
_{11}^{(1)}+\sigma _{22}^{(1)})-d_{31}^{(2)}\sigma _{31}^{(1)}, \\
&&Q^{(1)}=0,\,\,\,\,\,\,\,\,Q^{(2)}=-i\frac{\partial }{\partial t_{1}}\sigma
_{32}^{(1)}-\Omega _{p}^{(1)}\sigma _{21}^{(1)\ast }, \\
&&Q^{(3)}=-i\frac{\partial }{\partial t_{1}}\sigma _{32}^{(2)}-\Omega
_{p}^{(1)}\sigma _{21}^{(2)\ast }-\Omega _{p}^{(2)}\sigma _{21}^{(1)\ast
}--d_{32}^{(2)}\sigma _{32}^{(1)}, \\
&&Y^{(1)}=0,\,\,\,\,\,\,\,\,Y^{(2)}=-i\frac{\partial }{\partial t_{1}}\sigma
_{11}^{(1)}-\Omega _{p}^{(1)\ast }\sigma _{31}^{(1)}+\Omega _{p}^{(1)}\sigma
_{31}^{(1)\ast }, \\
&&Y^{(3)}=-i\frac{\partial }{\partial t_{1}}\sigma _{11}^{(2)}-\Omega
_{p}^{(1)\ast }\sigma _{31}^{(2)}+\Omega _{p}^{(1)}\sigma _{31}^{(2)\ast
}-\Omega _{p}^{(2)\ast }\sigma _{31}^{(1)}+\Omega _{p}^{(2)}\sigma
_{31}^{(1)\ast }, \\
&&Z^{(1)}=0,\,\,\,\,\,\,\,\,Z^{(2)}=-i\frac{\partial }{\partial
t_{1}}\sigma _{22}^{(1)}, \,\,\,\,\,\,\,\,Z^{(3)}=-i\frac{\partial
}{\partial t_{1}}\sigma _{22}^{(2)}.
\end{eqnarray}
\end{subequations}

It is convenient to express Eq.~(\ref{App1}) in the following form~\cite{huang}

\begin{subequations}
\label{App2}
\begin{eqnarray}
&&\hat{L}\Omega_{p}^{(m)}=S^{(m)}, \\
&&\sigma_{31}^{(m)}=\frac{1}{\kappa_{13}}\left[ M^{(m)}-i\left( \frac{%
\partial }{\partial z_{0}}+\frac{1}{c}\frac{\partial }{\partial t_{0}}%
\right) \Omega_{p}^{(m)}\right] , \\
&&\sigma_{21}^{(m)}=\left( i\frac{\partial }{\partial
t_{0}}+d_{21}^{(0)}\right)^{-1}\left[ N^{(m)}-\Omega_{c}^{\ast
}\sigma_{31}^{(m)}\right] , \\
&&\sigma_{11}^{(m)}=\left[ i\left( \frac{\partial }{\partial
t_{0}}+\Gamma_{13}\right)\hat{L}_{2}-i\Gamma_{13}\hat{L}_{1}\right]^{-1}\left\{
\hat{L}_{2}Y^{(m)}-i\Gamma_{13}\left[ Z^{(m)}\right. \right.  \notag \\
&&\quad\quad \left. \left. -\Omega_{c}^{\ast }\left( i\frac{\partial
}{\partial t_{0}}+d_{32}^{(0)}\right)^{-1}Q^{(m)}+\Omega_{c}\left(
-i\frac{\partial }{
\partial t_{0}}+d_{32}^{(0)\ast }\right)^{-1}Q^{(m)\ast }\right] \right\},
\\
&&\sigma_{22}^{(m)}=\frac{1}{\Gamma_{13}}\left[
-iY^{(m)}-\left(\frac{\partial }{\partial t_{0}}+\Gamma
_{13}\right)\sigma
_{11}^{(m)}\right] , \\
&&\sigma_{32}^{(m)}=\left( i\frac{\partial }{\partial
t_{0}}+d_{32}^{(0)}\right)^{-1}\left[ Q^{(m)}-\Omega_{c}(\sigma
_{11}^{(m)}+2\sigma_{22}^{(m)})\right] ,
\end{eqnarray}
\end{subequations}%
where operators $\hat{L}$, $\hat{L}_{1}$, $\hat{L}_{2}$ and
$S^{(m)}$ are given as
\begin{subequations}
\begin{eqnarray}  \label{L}
&&\hat{L}= i\left(\frac{\partial }{\partial
z_{0}}+\frac{1}{c}\frac{\partial
}{\partial t_{0}}\right)+\kappa _{13}\left(i\frac{\partial }{\partial t_{0}}%
+d_{21}\right)\left[|\Omega _{c}|^{2}-\left(i\frac{\partial }{\partial t_{0}}+d_{21}\right)\left(i%
\frac{\partial }{\partial t_{0}}+d_{31}\right)\right]^{-1},  \nonumber\\
&&S^{(m)}= M^{(m)}-\kappa _{13}\left[|\Omega_{c}|^{2}-\left(i\frac{%
\partial }{\partial t_{0}}+d_{21}\right)\left(i\frac{\partial }{\partial t_{0}}%
+d_{31}\right)\right]^{-1}\left[\Omega _{c}N^{(m)}-\left(i\frac{\partial }{\partial t_{0}}%
+d_{21}\right)P^{(m)}\right], \nonumber\\
&&\hat{L}_{1}= i\Gamma _{23}-|\Omega _{c}|^{2}\left[\left(i\frac{\partial }{%
\partial t_{0}}+d_{32}^{(0)}\right)^{-1}-\left(-i\frac{\partial }{\partial t_{0}}%
+d_{32}^{{(0)} \ast}\right)^{-1}\right],  \nonumber\\
&&\hat{L}_{2}= i\left(\frac{\partial }{\partial t_{0}}
+\Gamma_{23}\right)-2|\Omega _{c}|^{2}\left[\left(i\frac{\partial }{\partial t_{0}}%
+d_{32}^{(0)}\right)^{-1}-\left(-i\frac{\partial }{\partial
t_{0}}+d_{32}^{{(0)}\ast}\right)^{-1}\right].  \nonumber
\end{eqnarray}
\end{subequations}%
Equation (\ref{App2}) can be solved order by order as shown in
the main text.




\begin{references}
\bibitem{note1}The frequency and wave number of the probe field are given by
$\omega_p+\omega$ and $k_p+K(\omega),$ respectively. Thus
$\omega=0$ corresponds to the center frequency of the probe
field.
\end{references}
\end{document}